\documentclass[preprint,aps,floatfix,showpacs,
preprintnumbers]{revtex4}
\usepackage{amssymb}
\usepackage{epsfig}
\usepackage{graphicx}
\usepackage{dcolumn}
\usepackage{bm}
\begin{document}

\title{Electronic Control and Readout of Qubit States
in Solid State
Quantum Computing Systems}
\author{A.~G.~Petukhov$^{1}$\footnote{Email:Andre.Petukhov@sdsmt.edu},
V.~V.~Osipov$^{2,3}$\footnote{Email:osipov@email.arc.nasa.gov},
and V.~N.~Smelyanskiy$^{2}$\footnote{Email:Vadim.N.Smelyanskiy@nasa.gov}}
\affiliation
{$^{1}$ Physics Department, South Dakota School of
Mines and Technology, Rapid City, SD 57701\\
$^{2}$NASA Ames Research Center, Moffett Field, CA 94035\\
$^{3}$Mission Critical Technologies, Inc.,
El Segundo, CA 90245}

\date{\today}

\begin{abstract}

We demonstrate that an $n^+/i/n^+$
junction is the most suitable candidate for electronic control and
readout of qubit states in quantum computing systems based on shallow
impurities. 
The signature of this system is that the  $n^+-$regions serve as
metallic electrodes separated form the $i-$region by a
self-induced barrier (internal workfunction). The $n^+/i/n^+$
system mimics the properties of a metal-vacuum-metal junction with
the qubit (impurity atom) placed in a ``vacuum'' $i$-region
between two ``metallic'' $n^+$ electrodes. We will show that the
self-induced barrier exists in a sufficiently wide range of the
concentration of dopants in the $n^+$-semiconductor (e.g. up to
$10^{21}$ cm$^{-3}$ for Si) and its height can be  controlled by
tuning the doping level. A  shallow donor placed in a vacuum
$i$-region will be populated with one electron in equilibrium. In
the case of Li donor in Si  the $n^+$-electrodes will be used for
a precision placement of the Li atom during the growth process;
for voltage control and manipulation of the qubit states; and for
a qubit readout by means of the optically stimulated resonant
tunnelling. Another important feature of our system is that the
qubit states (first two lowest energy levels of Li in Si) are
separated by an energy gap from a continuum of the  many-body
states of the controlling electrodes. 

\end{abstract}

\maketitle

\noindent {\bf Introduction}. Various systems have been proposed
during last decade as potential candidates for solid state quantum
computing (QC). All these systems are based on control of either
localized spins or localized charged states in semiconductors. In
particular, it was proposed to encode qubits on: nuclear spins of
$^{31}$P donors \cite{kane1998a,skinner2003a} or $^{29}$Si
isotopes \cite{ladd2002a} in Si matrix; spins of electrons
localized in semiconductor quantum dots (QDs)
\cite{loss1998a,imamoglu1999a,vrijen2000a,sherwin1999a,chen2000a,stievater2001a,brown2002a};
and on charged states of shallow impurities in Si
\cite{golding2003a,smelyanskiy2005a}. For all these cases the
distance $d$ between neighboring qubits ranges from 5nm
\cite{kane1998a,skinner2003a,ladd2002a} to 100nm
\cite{golding2003a,smelyanskiy2005a} and electronic control must
be used for a readout and for one- and two-qubit gates. It is
clear, that the distance $a$ between an individual qubit and its
controlling electrode has to be smaller than the interqubit
separation $d$ to ensure a proper space resolution of the qubits.
Thus the distance $a$ must be in the range $a\lesssim $~5nm
(nuclear spins \cite{kane1998a,skinner2003a,ladd2002a}) and
$a\sim$~20nm (shallow impurities in Si
\cite{smelyanskiy2005a,golding2003a}).

\begin{figure}[tbh]
\includegraphics[width=0.9\linewidth,clip=true]{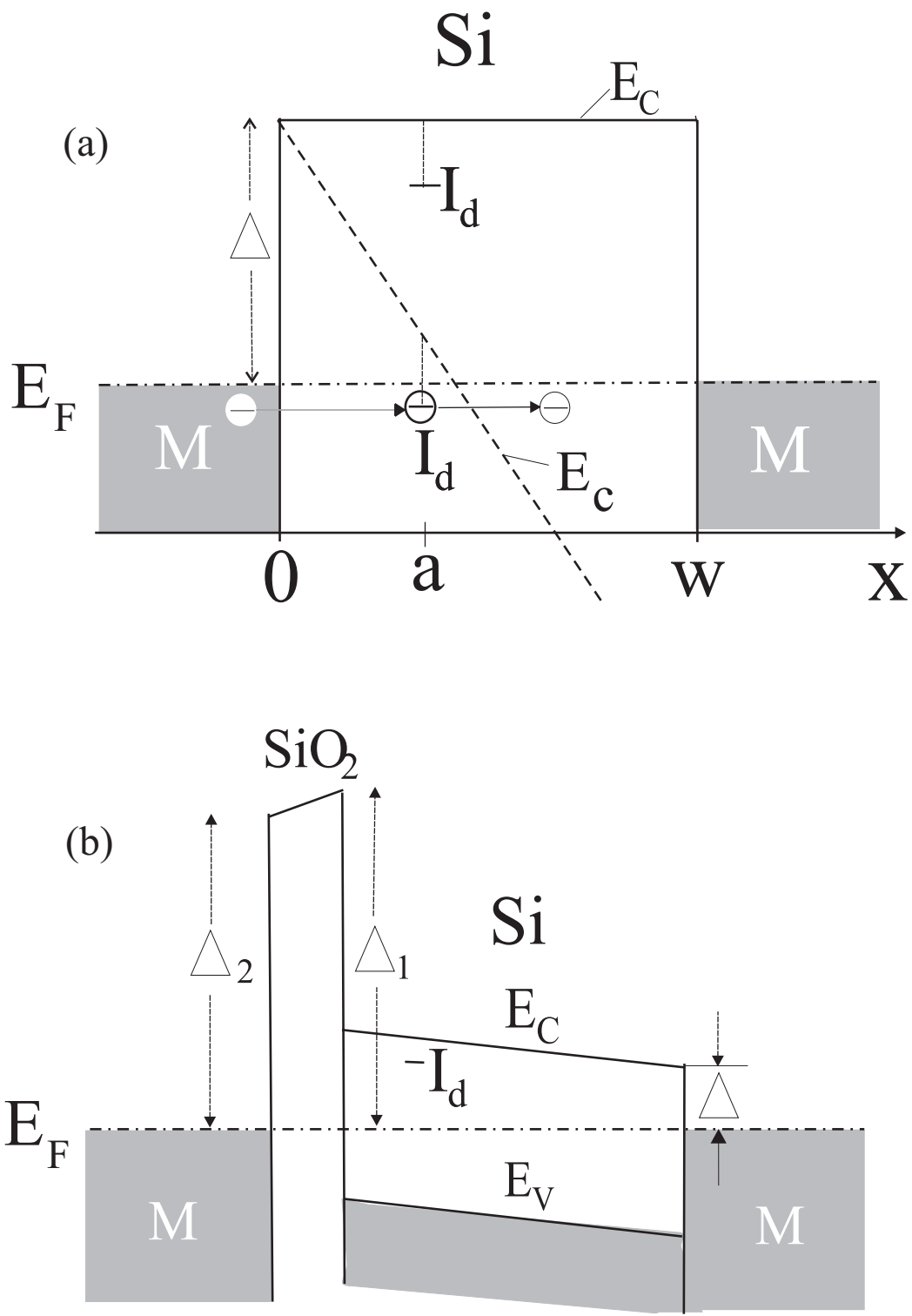}
\caption{Schematic band diagrams of Metal-Si-Metal (a), and
Metal-SiO$_2$-Si-Metal junctions (b).} \label{MS-MIS}
\end{figure}

It seems almost obvious that one can use metal electrodes for
control and readout of the solid state qubits. As it has been
suggested in most of the solid state QC proposals
\cite{kane1998a,skinner2003a}-\cite{golding2003a}, the thin-film
metal electrodes have to be deposited directly on a Si surface or
on a thin insulating SiO$_{2}$ layer grown on this surface. A
careful analysis of this idea reveals that a single shallow donor in
an undoped semiconductor placed in a proximity ($a\sim$ 5-20nm) of
an interface with the metal will  be ionized. It simply means that
the qubit {\em cannot be formed}.  This conclusion is in agreement
with the fact the localized electrons on shallow impurities placed
in  vicinity of  metal
electrodes have not been  observed in  QC experiments. Indeed, it
is well known that a high potential barrier (a Schottky barrier)
is always formed at  metal-semiconductor (MS) or
metal-insulator-semiconductor (MIS) interfaces  \cite{sze1981a}.
The barrier height is determined by a large density of the
interfacial states that pin the Fermi level $E_F$ at energy
$\Delta\simeq 0.5$-$0.8$~eV below the conduction band edge
$E_c(0)$ at the interface. The pinning occurs for Si and GaAs in
contacts with almost  all metals and for doping concentrations
$N\lesssim 10^{16}$ cm$^{-3}$ the barrier width $l_d$ is in the
range $l_{d}\gtrsim 100$ nm \cite{sze1981a}. However an  extremely
pure ($i$-)S layer with defect density $N\lesssim 10^{10}$
cm$^{-3}$ has to be used for the solid state QC. For these
concentrations the barrier width $l_d$ is in the range of $\mu$m
or more, and inside $i$-S layer the conduction band edge $E_c(z)$
is practically constant at the distances to the interface
corresponding to the qubit location, $z \le a\sim$~20 nm.
Therefore the potential of the $i$-S layer between two control
metal electrodes will be  flat
(see Fig.~\ref{MS-MIS}(a)%
). The barrier height is in the range 0.5-0.8 eV and all shallow
donors with ionization energy $I_d \lesssim$~100 meV will be ionized.

One can apply an electric field to push the impurity level below
the Fermi energy of the left electrode (see Fig.~\ref{MS-MIS}(a)).
However the required electric field,  $E\simeq \Delta /ae\gtrsim
3\times 10^{5}$V/cm ($e$ is the elementary charge), is so strong
that the impurity will be ionized due to the Zener tunnelling
breakdown into the right contact \cite{sze1981a}.

The potential barrier can be efficiently decreased if
a very thin
heavily doped $n^{+}$-S
layer with donor concentration $N_{d}^{+}\sim 10^{20}$ cm$^{-3}$
is formed between M
and $i$-S layers \cite{sze1981a}.
The required conditions will be fulfilled if
 $E_F<E_{c}(a)<E_F+I_d$  in $i$-semiconductor layer
at a distance $a$ (cf. Fig.~\ref{MS-MIS}(a)). Unfortunately, these
extremely stringent conditions cannot be satisfied simultaneously
for a sufficiently large interface plane and a sequence of donor
qubits located near it,  because of the large fluctuations of the
Schottky barrier $\Delta $ and the number of donors
 $N_{d}^{+}$ in the thin $n^+$ layer
\cite{das1989a,palm1993a,tung1992a,zheng1999a}. Similar
overwhelming obstacles also arise in MIS structures with thin
insulating layers (Fig.~\ref{MS-MIS}(b)). We note that the barrier
height $\Delta _{1}$ at SiO$_{2}$-Si interface is about 4 eV
\cite{sze1981a} and therefore it is impossible to push the donor
level below $E_F$ with any realistic electric fields.

The purpose of this paper is to demonstrate that an $n^+/i/n^+$
junction is the most suitable candidate for electronic control and
readout of qubit states in quantum computing systems based on
shallow
impurities. 
The signature of this system is that the  $n^+-$regions serve as
metallic electrodes separated form the $i-$region by a
self-induced barrier (internal workfunction). The $n^+/i/n^+$
system mimics the properties of a metal-vacuum-metal junction with
the qubit (impurity atom) placed in a ``vacuum'' $i$-region
between two ``metallic'' $n^+$ electrodes. As we will see below,
the self-induced barrier exists in a sufficiently wide range of
the concentration of dopants in the $n^+$-semiconductor (e.g. up
to $10^{20}$ cm$^{-3}$ for Si) and its height can be  controlled
by tuning the doping level. A  shallow donor placed in a vacuum
$i$-region will have its ground state energy below the Fermi level
and will be populated with one electron in equilibrium.

Finally, we note  that for a qubit based on Li donor in Si
\cite{smelyanskiy2005a} the same $n^+$-electrodes can be be used
for a precision placement of the electro-migrating Li ions during
the growth process; for voltage control and manipulation of the
qubit states in Li at low temperatures; and for a qubit readout by
means of the optically stimulated resonant tunnelling.

\vspace{0.2in}

\begin{figure}[tbh]
\includegraphics[width=0.9\linewidth,clip=true]{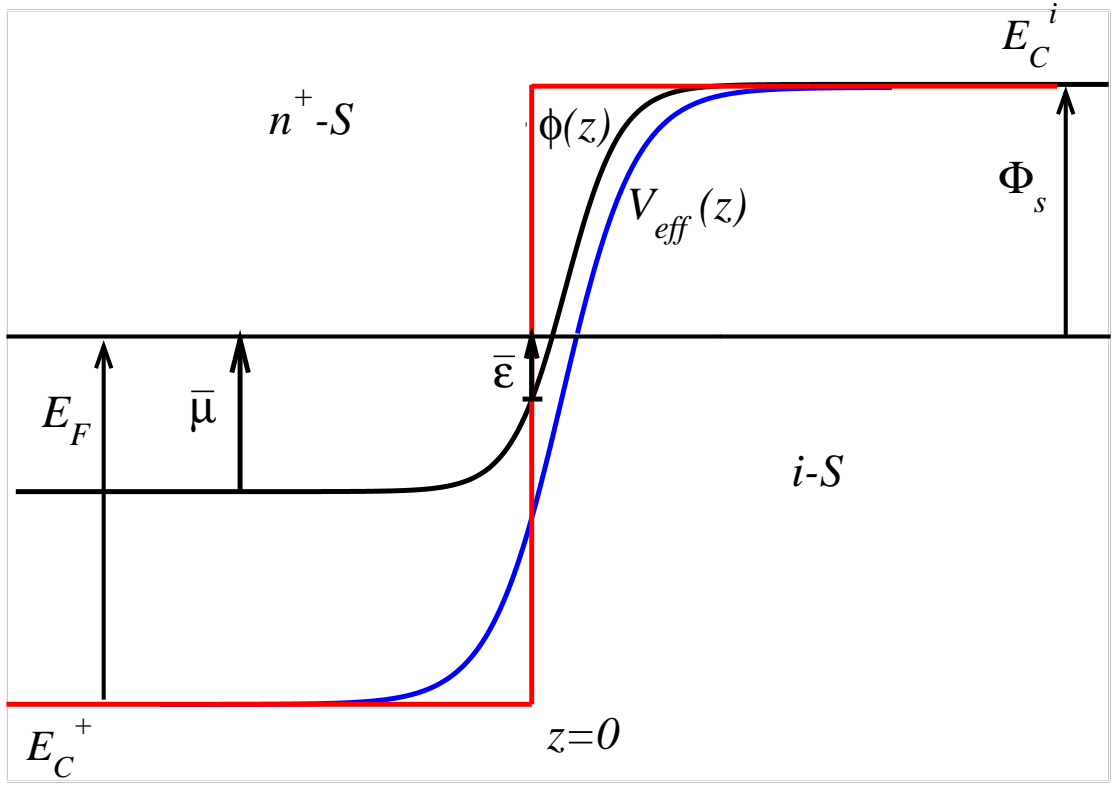}
\caption{Band diagram of $n^+/i$ junction and schematic behavior
of $\phi(x)$ and $V_{\rm eff}(z)$} \label{potential}
\end{figure}

 \noindent
{\bf LDA treatment of doped semiconductor $^+$-i structure}. Let
us consider formation of a barrier $\Phi_s$ at a $n^+/i$
interface. We will assume that the donor concentration in
$n^+$-region exceeds the threshold of the metal-insulator phase
transition. If we replace spatial distribution of ionized donors
with positive uniform charge and consider a single-valley
parabolic-band $n$-type semiconductor (e.g. GaAs) the problem can
be mapped onto the problem of a metal-vacuum interface
\cite{bardeen1936a,smith1969a,lang1971a}. The barrier of an
$n^+/i$ junction or, equivalently, the workfunction of a simple
metal can be expressed in the form known as Budd-Vannimenus
theorem \cite{budd1973a,mahan1974a}:
\begin{equation}
\Phi_s=
\phi(\infty)-\phi(0)-\bar\varepsilon.\label{mahan}
\end{equation}
The meaning of Eq.~(\ref{mahan}) is
illustrated in Fig.~\ref{potential} where a schematic band diagram
of the $n^+/i$-junction is shown.
The energies $E_C^+$ and $E_C^i$ are the conduction band edges of
$n^+$ and $i$-semiconductors respectively, $V_{\rm eff}(z)$ is the
exact one-electron quasi-particle potential, and $\phi(z)$ is the
exact electrostatic (Hartree) energy of the electron related to
the electron density $n(z)$  through the Poisson's equation:
\begin{equation}
\frac{d^2\phi}{dz^2} = \frac{4\pi N e^2}{\varkappa}
\left(\Theta(-z)-\frac{n}{N}\right),
\label{Poisson}
\end{equation}
where $\Theta(y)$ is a step function. The quantities
$\bar\mu$ and $\bar\varepsilon$ are
the exact chemical potential
many-body effects included \cite{lang1971a} and
the exact total energy
per one electron in the bulk respectively. All these
quantities are well known within the local density approximation
(LDA).
In particular,
\begin{equation}
\label{epsilon} \bar\varepsilon=
\frac{3}{5}E_F+\frac{3}{4}E_{ex}+E_c=
\left(\frac{2.21}{r_s^2}-\frac{0.916}{r_s}-\frac{0.88}{r_s+7.8}
\right)Ry^*,
\end{equation}
where we introduced the effective Rydberg:
\begin{equation}
\label{Rydberg}
Ry^*=\frac{e^2}{2\varkappa a_0}=\frac{\hbar^2}{2m^*a_0^2},
\end{equation}
$\varkappa$ is the dielectric constant, $m^*$ is the effective mass, and
$a_0$ is the effective Bohr radius:
\begin{equation}
\label{bohr} a_0=\frac{\varkappa\hbar^2}{m^*e^2}.
\end{equation}
The parameter $r_s$ is related to the bulk electron concentration
$n_0$ in the $n^+$ region as follows:
\begin{equation}
\label{rs}
r_s=\left(\frac{3}{4\pi n_0a_0^3}\right)^{1/3}=
\left(\frac{3}{4\pi Na_0^3}\right)^{1/3},
\end{equation}
where $N$ is the donor concentration and we assume that
all the donors are ionized, i.e. $n_0=N$.

Since $\bar\varepsilon$ is known the problem of finding the
barrier $\Phi_s$ is reduced to calculation of the potential
difference $\phi(\infty)-\phi(0)$ caused by the interfacial dipole
layer. The latter can be expressed as:
\begin{equation}
\label{dipole}
\phi(\infty)-\phi(0)=
\frac{4\pi e^2Q}{\varkappa k_F}
\frac{\int_0^1g(\epsilon)d\epsilon\int_0^\infty \zeta\left
|\psi_\epsilon(\zeta)\right|^2d\zeta}
{\int_0^1g(\epsilon)d\epsilon\int_0^\infty \left
|\psi_\epsilon(\zeta)\right|^2d\zeta},
\end{equation}
where we introduced dimensionless energy $\epsilon=[E-V_{\rm
eff}(-\infty)]/E_F$, and coordinate
$\zeta=k_Fz\equiv(3\pi^2N)^{1/3}z$. Here $Q$ is the total number
of electrons per unit area in the  half-space $\zeta < 0$, and
\begin{equation}
g(\epsilon)=\frac{3}{4}\frac{1-\epsilon}{\sqrt{\epsilon}}
\end{equation}
is a dimensionless effective density of the occupied
states for electrons moving
perpendicular to the interface.

Eq.~(\ref{dipole}) is exact and follows from the properties of the
1-dimensional Poisson's equation.
The internal workfunction $\Phi_s$ can be calculated exactly
provided
that we know the eigenfunctions
of the Kohn-Sham equations  $\psi_\epsilon(z)$:
\begin{equation}
\label{Kohn-Sham}
\left[-\frac{\hbar^2}{2m^*}\frac{d^2}{dz^2}+V_{\rm
eff}(z)\right]\psi(z)= E\psi(z)
\end{equation}
with
\begin{equation}
V_{\rm eff}(z)=\phi(z)+E_{ex}(n(z))+E_c^\prime(n(z))\equiv
\phi(z)-\frac{1.22}{r_z}Ry^*-0.88 \frac{7.8+4 r_z/3}{(7.8+
r_z)^2}Ry^*,
\end{equation}
where $r_z=(4\pi n(z)a_0^3/3)^{-1/3}$, electrostatic energy
$\phi(z)$ is related to $n(z)$ via Eq.~(\ref{Poisson}), and
\begin{equation}
\label{density} n(z)=N\int_0^1g(\epsilon)|\psi_\epsilon
(z)|^2d\epsilon.
\end{equation}

The detailed numerical solution of
Eqs.~(\ref{Kohn-Sham})-(\ref{density}) will be published
elsewhere. Here we will estimate the interface dipole term using
approximate eigenfunctions of the one-electron Schrodinger
equation with the potential in the form of a rectangular step of
the height $E_F+\Phi_s$ (shown in red in Fig.~\ref{potential}):
\begin{equation}
\label{psi}
|\psi_\epsilon(\zeta)|^2=\frac{2\epsilon}{1+\xi_0}
\exp[-2\sqrt{1+\xi_0-\epsilon}\cdot \zeta],
\end{equation}
where $\xi_0=\Phi_s/E_F$. The wavefunctions in Eq.~(\ref{psi})
capture the most important feature of the exact solution -
exponential asymptotic decay far from the interface
\cite{levy1984a}. An estimate for the value of the parameter $Q$
can be obtained from the  Thomas-Fermi equation inside
$n^+$-region  ($z<0$). This yields:
\begin{equation}
\label{Q}
Q=0.088r_s^{-5/2}a_0^{-2}.
\end{equation}
Substituting Eqs.~(\ref{psi}) and~(\ref{Q}) in Eq.~(\ref{dipole})
we  obtain:
\begin{equation}
\label{phis}
\Phi_s\equiv\frac{3.68}{r_s^2}\xi_0=
\frac{0.982}{r_s^{3/2}}f(\xi_0)-
\frac{2.21}{r_s^2}+\frac{0.916}{r_s}+
\frac{0.88}{r_s+7.8},
\end{equation}
where
\begin{equation}
f(y)=\frac{\pi}{2}
\frac{1+3y-3y\sqrt{1+y}\cdot\coth^{-1}\sqrt{1+y}} {\sqrt{y}(1+3
y)+(1+y)(1-3y)\cot^{-1}\sqrt{y}}.
\end{equation}
\begin{figure}[tbh]
\includegraphics[width=0.8\linewidth,clip=true]{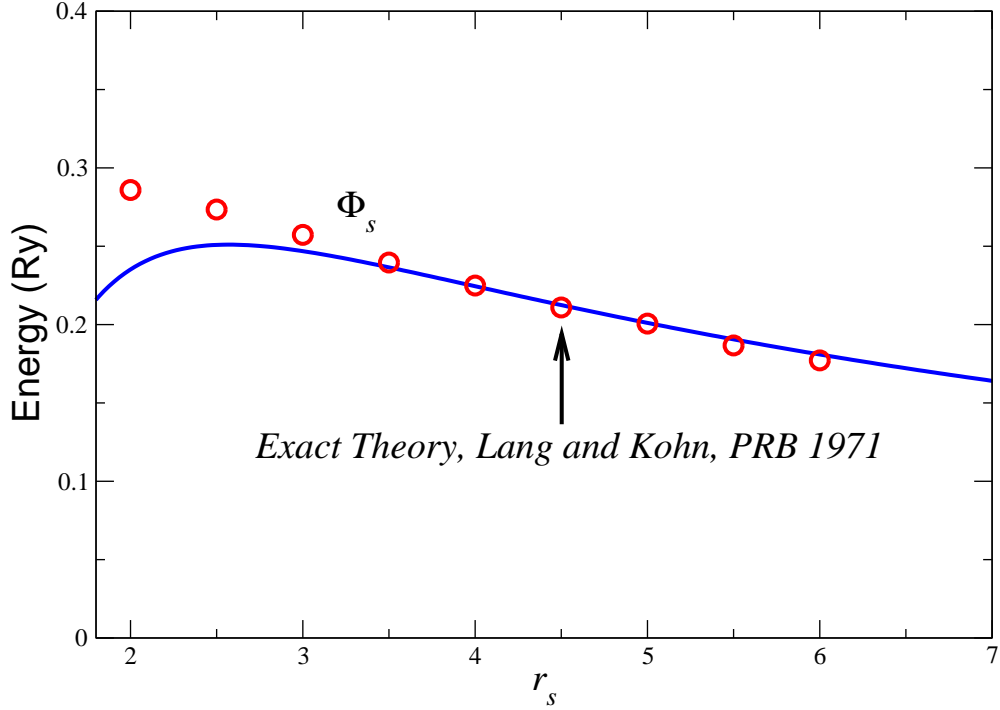}
\caption{Workfunction of a jellium metal at different densities:
blue curve - Eq.~(\protect{\ref{phis}}), circles - exact
theory \protect{\cite{lang1971a}}}
\label{workfunction5}
\end{figure}
The results of the application of Eq.~(\ref{phis}) are shown in
Fig.~\ref{workfunction5} and compared with the exact numerical
results by Lang and Kohn \cite{lang1971a}. We see that our
estimate is in good agreement with the exact theory for
$r_s\gtrsim$~2.5. As we will see below this region of electron
densities is the most relevant one for quantum computing purposes
in  $n^+/i/n^+$ Si system.

\begin{figure}[t]
\includegraphics[width=0.9\linewidth,clip=true]{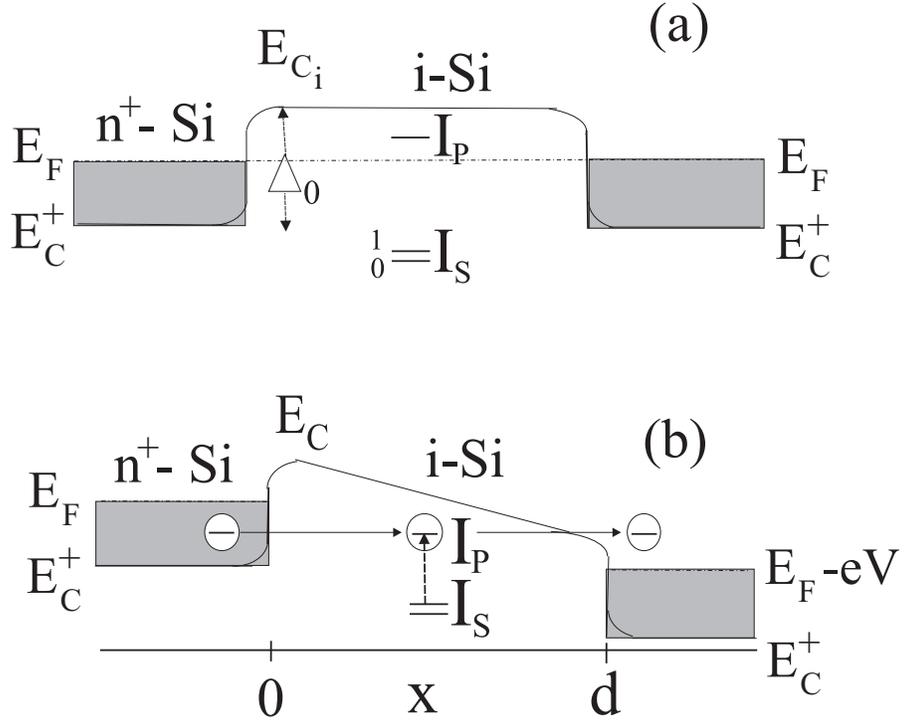}
\caption{Schematic band
 diagrams of n$^{+}$-$i$ structure in equilibrium (a) 
and at a bias voltage $V$ (b).}
\label{RT}
\end{figure}

 In order to apply our results to Si we have to take into
account multi-valley character of the kinetic energy operator. If
we neglect the effective mass anisotropy the main effect will
consist in replacing parameter $k_F$ with its renormalized value
$(3\pi^2N/g_v)^{1/3}$, where $g_v$ is the number of valleys in the
conduction band. The analog of Eq.~(\ref{phis}) reads:
\begin{equation}
\label{phis2}
\Phi_s\equiv\frac{3.68}{g_v^{2/3}r_s^2}\xi_0=
\frac{0.982}{r_s^{3/2}}f(\xi_0)-
\frac{2.21}{g_v^{2/3}r_s^2}+\frac{0.916}{g_v^{1/3}r_s},
\end{equation}
\noindent
\begin{figure}[t]
\includegraphics[width=0.8\linewidth,clip=true]{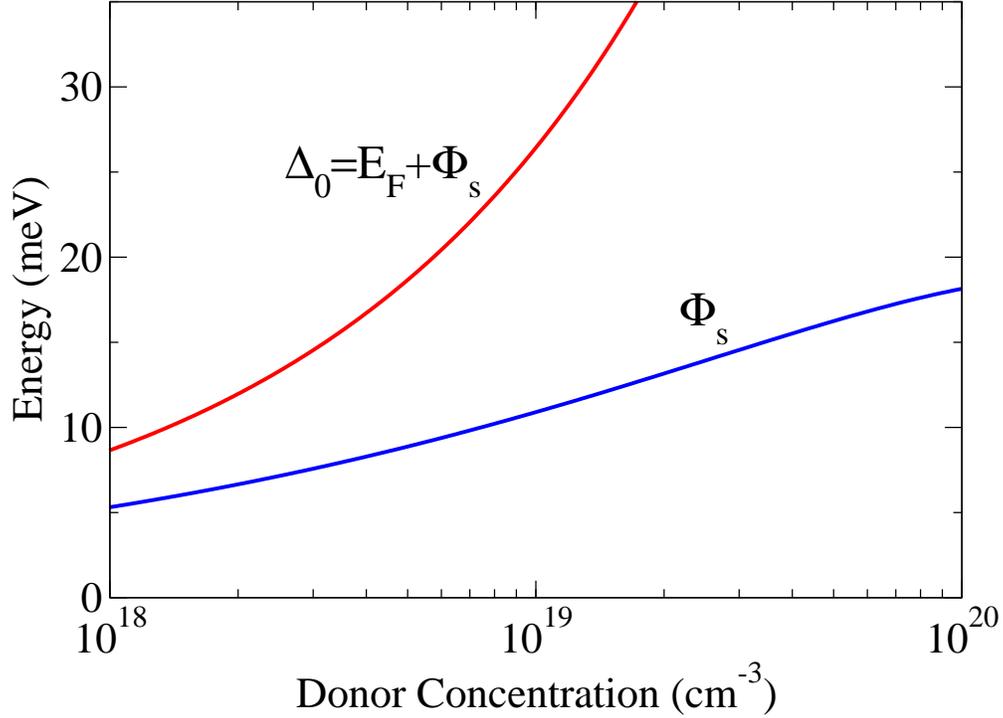}
\caption{Barrier $\Phi_s$ of Si $n^+/i$ junctions at different
donor concentrations (blue), Eq.~(\protect{\ref{phis2}}), the
total conduction band offset $\Delta_0=E_F+\Phi_s$ is also shown
in red.} \label{workfunction7}
\end{figure}
\noindent
 where the number of valleys $g_v$=6 for
Si. Here we have omitted insignificant correlation energy.  The
results of the calculations of $\Phi_s$ by means of
Eq.~(\ref{phis2})
 are shown in Fig.~\ref{workfunction7}
We used $Ry^*$ = 34.5 meV, and $a_0$ = 18.3~\AA
\cite{perrera1995a}. It can be seen from Fig.~\ref{workfunction7}
that for the purposes of quantum computing on long-lived donor
states of Li in Si \cite{smelyanskiy2005a} the electron
concentration in $n^+$ electrodes must be in the range between
2.55$\cdot$10$^{18}$~cm$^{-3}$ ($g_v^{1/3}r_s=4.5$) and
1.5$\cdot$10$^{19}$~cm$^{-3}$ ($g_v^{1/3}r_s=2.5$). The lower
limit is the metal-insulator transition threshold for $n^+$ Si
\cite{shklovskii1984a,perrera1995a} while the upper limit
corresponds to the point where the total conduction band offset
$\Delta_0=E_F+\Phi_s$ between $n^+$- and $i$-layers reaches 1
Ry$^*$. A schematic energy diagram corresponding to this range of
concentrations is shown in Fig.~\ref{RT}~(a). It can be seen that
the first two lowest energy levels of Li donor in Si are separated
by an energy gap from a continuum of the many-body states in the
$n^+$-electrodes. Thus our qubit is protected from a dephasing
caused by dynamic fluctuations of the electron density in the
contacts.

\vspace{0.2in}

 \noindent {\bf Qubit state readout via
photo-assisted resonant tunnelling}. The $n^{+}$-$i$-$n^{+}$ or
$p^{+}$-$i$-$p^{+}$ structures can be used for readout the
electron states in systems considered in Refs.~\cite{golding2003a,
smelyanskiy2005a}. Indeed, and individual qubit in these systems
is encoded on the ground and long-lived first excited state of a
shallow donor or acceptor. According to estimates of Ref.
\cite{smelyanskiy2005a} for one of those systems, Li donor in Si,
the
lifetime of the first excited state 1s,B$%
_{2}$ (further denoted as $| 1\rangle$) can be up to 1 second  if
the energy separation between this state and the ground state
1s,A$_{1}$  (further denoted as $| 0\rangle$) is $~\sim $ 0.06
meV. The ionization energy of the ground state of Li in Si,
$I_{0}$=1~Ry$^*$=34.5~meV.
The excited $%
2p_{0}^{-}$-state has the ionization energy $I_{p}$=11.5meV and
the lifetime $\tau _{p}\sim $10$^{-8}$s \cite{pavlov2000a}. The
readout can be realized when the lower levels $|0 \rangle$ and $|
1 \rangle$
lie below $%
E_{c}^{+}$ and the $2p_{0}^{-}$-level is higher than $E_{c}^{+}$
(Fig.~\ref{RT}(b)). This situation corresponds to doping
concentration $N\lesssim 10^{19}$~cm$^{-3}$. In this case the
lower levels of the qubit lie below the continuum spectrum of the
contacts  in equilibrium and one of them is populated by the
electron which does not interact with electrons in conducting
$n^{+}$ layers (Fig.~\ref{RT}(b)). When bias voltage $V\sim$30~meV
is applied the $2p_{0}^{-}$-level moves down below the Fermi level
of the left electrode
and all  Li levels are shifted under the action of electric field $E=V/d$%
$\sim$~4KV/cm (Fig.~\ref{RT}(c)). Radiation absorption between lower Li states
$| 0\rangle$ or $| 1 \rangle$
and the state 2p$_{0}$ depends on the polarization of the infrared
light. Thus, selecting the radiation by frequency and polarization
we can excite only the state $| 0\rangle$ or $| 1 \rangle$
depending on the polarization  and detect the tunnelling current
through the $2p_{0}^{-}$-level of the ionized impurity
(Fig.~\ref{RT}(c)).
The tunnelling frequency can be estimated as%
\[
\omega =(E_F/\hbar )\exp (-2d/3l_{0}),
\]%
where $l_{0}=(\hbar ^{2}/2m^*\Phi_{s})^{1/2}$ is the tunnelling
length. For typical values $\Phi _{s}$=8meV, $E_F \simeq $20meV
and $d$=25nm we obtain $\omega >$10$^{10}$ Hz, i.e. $\omega \tau
_{p}\gg 1$. It means that many electrons will tunnel through the
excited $2p_{0}^{-}$-level before the recombination process
occurs, the donor becomes neutral, and the current terminates.
Thus a single-qubit measurement with the current signal much
stronger than that of the single-electron measurement can be
realized. The tunnelling current depends dramatically on Li
position, namely it has a sharp maximum when the Li atom is placed
equidistantly between the
electrodes \cite%
{xu1995a,osipov2004a}. This effect can be used for {\it in situ}
placement control of Li atoms during the growth process at high
temperatures, when the highly-mobile Li ions are guided to their
permanent final positions  by an external electric field.

The effect of photo-stimulated resonant tunnelling can be verified
independently in the experiments on $\delta$-doped Li layers
placed between two $n^+$-electrodes. These structures are similar
to those studied in Ref.~\cite{caro2003a}. Consider kinetics of
the resonant current immediately after the photoionization occurs.
For simplicity we assume that the tunnelling frequency is
relatively small, $\omega/W_{dec}\ll 1$, where $W_{dec}\geq
\tau_{p}^{-1}$ is a total decoherence width of the 2$p_{0}^{-}$
level. Then the rate equations for \emph{sequential} resonant
tunnelling read:
\begin{eqnarray}
dn_1/dt &=& \Gamma_l(1-n)-\Gamma_rn_1-Wn_1(1-n_0) \label{n1}\\
dn_0/dt &=& Wn_1(1-n_0).\label{n0}
\end{eqnarray}
Here $n_i$ are the populations of the levels $0$ and $1$,
$n=n_1+n_0$, $\Gamma_l$ and $\Gamma_r$ are inverse elastic
tunnelling lifetimes, and W is the inverse recombination time. The
first term in Eq.~(\ref{n1}) takes into account the Coulomb
blockade effect. The difference between
Eqs.~(\ref{n1}),~(\ref{n0}) and conventional equations
\cite{Dattabook} is that the resonant current is decaying via the
recombination channel.
If $\Gamma_{l(r)} \gg W$ the population $n_1$ quickly attains
quasi-equilibrium and
the tunnelling current can be expressed as:
\begin{equation}
I=\Gamma_l(1-n) \sim \frac{\Gamma_l\Gamma_r}{\Gamma_l+\Gamma_r}
\frac{1}{1+t/\tau},\label{current}
\end{equation}
where
\begin{equation}
\tau^{-1}=\frac{\Gamma_lW}{\Gamma_l+\Gamma_r}. \label{tau}
\end{equation}
Therefore, the photo-stimulated resonant tunnelling current
through the $\delta$-doped Li layer will be characterized by a
long-term decay kinetics $I\sim 1/t$. We note that the condition
for the sequential tunnelling consider above  is opposite to that
required for the qubit readout. In the letter case
$\omega_{t}/W_{dec}\gg 1$ and a coherent resonant tunnelling
regime needs to be considered.

\vspace{0.1in}

\noindent {\bf Conclusion} We proposed a device based on
 $n^+ - i - n^+$ heterostructures for electronic control and readout of 
 qubit encoded in
 two lowest states of a shallow donor of Li in Si.
Qubit state readout is based on the polarization-sensitive
photo-stimulated resonant tunnelling effect. A key feature of the
system is  that the qubit states (first two lowest energy levels
of a donor) are separated by a finite energy gap from a continuum
of the many-body states of the controlling electrodes. Therefore
the qubit is protected from a dephasing caused by dynamic
fluctuations of the electron density in the contacts. It is of
interest to apply  the proposed scheme for the case of qubits
based on orbital states of quantum dots as well as on shallow
acceptor states \cite{golding2003a} (in the later case
$p^{+}$-$i$-$p^{+}$ structures should be  used).

\newpage
\appendix
\section{Thomas-Fermi equation}
According to the Thomas-Fermi theory the electrons are considered as
classical particles obeying Fermi statistics.Then the
maximal energy of an electron near point $z$
reads:
\begin{equation}
\label{Emax}
E_{max}=\frac{p_{max}^2}{2m}+\phi(z), 
\end{equation}
where $\phi(z)=-e\varphi(z)$ is the electrostatic (Hartree) energy
of an electron in the field of all other electrons and external potential,
$e>0$  is the magnitude of the elementary charge, $\varphi$
is the electrostatic potential, and $p_{max}$
is the maximal momentum of the electron. The maximal energy of the
electron is nothing but a chemical potential at $T=0$. This chemical
potential must be constant in equilibrium:
\begin{equation}
\label{chempot}
E_{max}=const=E_F,
\end{equation}
where
\begin{equation}
\label{Fermi}
E_F=\frac{\hbar^2}{2m^*}(3\pi^2n_0)^{2/3}=\frac{\hbar^2}{2m^*}(3\pi^2N)^{2/3}=
\frac{\hbar^2k_F^2}{2m^*}.
\end{equation}
Here $n_0=N$ is the electron density in the bulk and $N$ is the
concentration of donors.
The maximal(Fermi) momentum can be related to the number density
of the electrons $n(z)$:
\begin{equation}
\label{pmax}
p_{max}=\hbar k_F[n(z)]=\hbar(3\pi^2n(z))^{1/3}
\end{equation}
Thus, from Eqs.~(\ref{Emax})-(\ref{pmax}) the relation between
electron density $n(z)$ and electrostatic energy $\phi(z)$ within
the Thomas-Fermi model is:
\begin{equation}
\label{n}
n=N\left(1-\frac{\phi}{E_F}\right)^{3/2}
\end{equation}

The Poisson equation for $\phi$ reads:
\begin{equation}
\frac{d^2\phi}{dz^2} = \frac{4\pi N e^2}{\varkappa}
\left(\Theta(-z)-\frac{n}{N}\right),
\label{Poisson2}
\end{equation}
Introducing dimensionless energy $u=1-\phi/E_F$,
from Eqs~(\ref{n})-(\ref{Poisson2}) we obtain the self-consistent,
dimensionless, Thomas-Fermi equation:
\begin{equation}
\label{u}
\frac{d^2 u}{d\zeta^2}=\frac{2}{3}
\left[u^{3/2}-\Theta(-\zeta)\right],
\end{equation}
Here we introduced dimensionless length $\zeta=z/L_{TF}$, where
$L_{TF}$ is the Thomas-Fermi screening length:
\begin{equation}
L_{TF}=\sqrt{\frac{E_F \varkappa }{6\pi N e^2}}
=\frac{1}{3}k_Fa_0^2r_s^{3/2}=
\frac{a_0}{2}\left(\frac{\pi}{3}\right)^{1/6}\frac{1}{(Na_0^3)^{1/6}}
\end{equation}
Here $a_0=\varkappa\hbar^2/m^*e^2$ is the Bohr radius.

The equation $u^{\prime\prime}=f(u)$ can be solved
as:
\begin{equation}
\left(\frac{du}{d\zeta}\right)^2=2\int f(u)du + Const
\end{equation}
Therefore, we obtain:
\begin{equation}
\label{du/dz}
\left(\frac{du}{d\zeta}\right)^2=
\frac{8}{15}u^{5/2}-\frac{4}{3}\Theta(-\zeta)
\left(u-\frac{3}{5}\right)
\end{equation}


It can be shown (\cite{mahan1974a}) that
$u(0)=3/5$. Therefore the solution of the Thomas-Fermi
equation for $\zeta>0$ reads:
\begin{equation}
\label{u_solution}
u=
\frac{3}{5}\frac{1}
{\left(1+\zeta/\zeta_0\right)^4},
\end{equation}
where $\zeta_0=(10\sqrt{15})^{1/2}$. Finally, from Eqs.~(\ref{n})
and (\ref{u_solution})
we obtain for $z>0$:
\begin{equation}
\label{n_solution}
\displaystyle
\frac{n}{N}=
\left(\frac{3}{5}\right)^{3/2}\frac{1}
{\left(1+\displaystyle{\frac{z}{(10\sqrt{15})^{1/2}L_{TF}}}\right)^6}
\end{equation}
Integrating Eq.~(\ref{n_solution}) from $0$ to $\infty$ we obtain
the total number of electrons per unit area in the upper half-space
(or holes in the lower half-space):
\begin{equation}
Q=\sqrt{2}\left(\frac{3}{5}\right)^{7/4}L_{TF}N=
\frac{0.088}{r_s^{5/2}a_0^{2}}
\end{equation}
and the prefactor in the expression for $\phi(\infty)-\phi(0)$:
\begin{equation}
\label{prefactor}
\frac{4\pi e^2 Q}{\varkappa k_F}=2\sqrt{2}\left(\frac{3}{5}\right)^{7/4}
r_s^{-3/2}Ry^*
\end{equation}

For a multi-valley semiconductor with isotropic spherical valleys
we have to replace Eq.~(\ref{pmax}) with
\begin{equation}
\label{pmaxm}
p_{max}=\hbar k_F[n(z)/g_v]=\hbar[3\pi^2n(z)/g_v]^{1/3},
\end{equation}
where $g_v$ is the number of valleys. Therefore, the screening length
and $Q$ will change accordingly, however Eq.~(\ref{prefactor}) will not.

\section{Electron Density}
To calculate the electron density $n(z)$ using LDA wavefunctions
we have to take into account the fact that we are dealing with the
wavefunction of a $3D$-semi-infinite metal:
\begin{equation}
\label{wavefunction}
\Psi(\vec r)=\frac{1}{\sqrt V}e^{\displaystyle{i\vec k_\parallel}\vec r}
\bar\psi_{k_z}(z)
\end{equation}
The quantum number $k_z$ describes the bulk states, i.e.
$\bar\psi_{k_z}(z)=A\psi_{k_z}(z)
\rightarrow A\sin(k_zz+\delta)$ as $z\rightarrow -\infty$.
Using Eq.~(\ref{wavefunction}) we can calculate the number density:
\begin{eqnarray}
n(z)&=&\frac{2}{V}\frac{V}{(2\pi)^3}\int_{-k_F}^{k_F}dk_z\int_0^{k_F}
2\pi k_\parallel dk_\parallel \Theta(k_F^2-k_\parallel^2-k_z^2)
\left|\bar\psi_{k_z}(z)\right|^2\\
&=&\frac{1}{2\pi^2}\int_0^{k_F}(k_F^2-k_z^2)
\left|\bar\psi_{k_z}(z)\right|^2dk_z
\end{eqnarray}
Introducing dimensionless energy $\epsilon=\hbar^2k_z^2/(2m^*E_F)$
we further obtain:
\begin{equation}
\label{nz2}
n(z)=\frac{1}{4\pi^2}\left(\frac{2m^*E_F}{\hbar^2}\right)^{3/2}
\int_0^1\frac{1-\epsilon}{\sqrt\epsilon}
\left|\bar\psi_\epsilon(z)\right|^2d\epsilon=
\frac{3}{4}N\int_0^1\frac{1-\epsilon}{\sqrt\epsilon}
\left|\bar\psi_\epsilon(z)\right|^2d\epsilon
\end{equation}
If we know $Q$ we can normalize the wavefunctions $\bar\psi_\epsilon(z)$
in the upper half-space:
\begin{equation}
\label{norm}
\left|\bar\psi_\epsilon(z)\right|^2=
\frac{Q\left|\psi_\epsilon(z)\right|^2}
{N\int_0^1d\epsilon g(\epsilon)\int_0^\infty\left|\psi_\epsilon(z)\right|^2dz}
\end{equation}
where
\begin{equation}
g(\epsilon)=\frac{3(1-\epsilon)}{4\sqrt{\epsilon}}
\end{equation}

\section{1D Poisson Equation}
The general solution of $1D$ Poisson equation (\ref{Poisson2}) reads:
\begin{equation}
\label{Poisson_solution}
\phi(z)-\phi(-\infty)=\frac{4\pi e^2}{\varkappa}
\int_{-\infty}^z(z^\prime-z)\rho(z^\prime)
dz^\prime,
\end{equation}
where $\rho(z)=n(z)-N\Theta(-z)$. If the system is electrically neutral,
then
\begin{equation}
\label{Poisson_solution2}
\phi(\infty)-\phi(-\infty)=\frac{4\pi e^2}{\varkappa}\int_{-\infty}^\infty
z^\prime\rho(z^\prime)
dz^\prime,
\end{equation}
and
\begin{equation}
\label{half-dipole}
\phi(\infty)-\phi(0)=\frac{4\pi e^2}{\varkappa}\int_{0}^\infty
z^\prime n(z^\prime)
dz^\prime.
\end{equation}
Substituting Eqs.~(\ref{nz2}) and (\ref{norm})
in Eq.~(\ref{half-dipole}) we obtain Eq.~(\ref{dipole})
of the paper.

\section{Modified Bardeen Treatment.}

J.~Bardeen, in his pioneering work of 1936 \cite{bardeen1936a}, 
calculated a moment of a
surface dipole layer using a model step-like potential 
barrier of infinite height. Her we will follow a similar
path but assume that the height of the  barrier
is finite. The value of the barrier will be determined
self-consistently. 

Let us approxiamte the exact self-consistent potential of a
semiinfinite jellium metal with a step
function:
\begin{equation}
V_{eff}(z)=V(-\infty)+(E_F+\Phi_s)\Theta[L(\xi_0)-z].
\end{equation}
Here, as previously, $\xi_0=\Phi_s/E_F$, and we assumed that the region
of the classical motion of electrons is extended beyond the metal surface
$z=0$. This assumption is  necessary to ensure elelctroneutrality
of the system. Integrating the charge densities inside and outside the
metal and imposing the electroneutrality condition we  find $L(\xi_0)$:
\begin{equation}
L(\xi_0)=\frac{q_1(\xi_0)-q_2(\xi_0)}{k_F},
\end{equation}
where
\begin{eqnarray}
{q_1}(\xi_0 )&=&\frac{3\pi }{8}-\frac{3}{16} \left[\frac{\sqrt{\xi_0 } (3+\xi_0 )}{1+\xi_0 }+
(3-\xi_0 ) \cot^{-1}(\sqrt{\xi_0 })\right]\\
q_2(\xi_0) &=& \frac{3}{16}\left[\frac{\sqrt{\xi_0 } (1+3 \xi_0 )}{1+\xi_0 }+(1-3 \xi_0 ) 
{\cot}^{-1}(\sqrt{\xi_0})\right]\\
\end{eqnarray}
The limiting value $L(\infty)=3\pi/8k_F$ was previously found
by Bardeen \cite{bardeen1936a}  
(see also Ref. \cite{newns1970a}).

\begin{figure}[tbh]
\includegraphics[width=0.8\linewidth,clip=true]{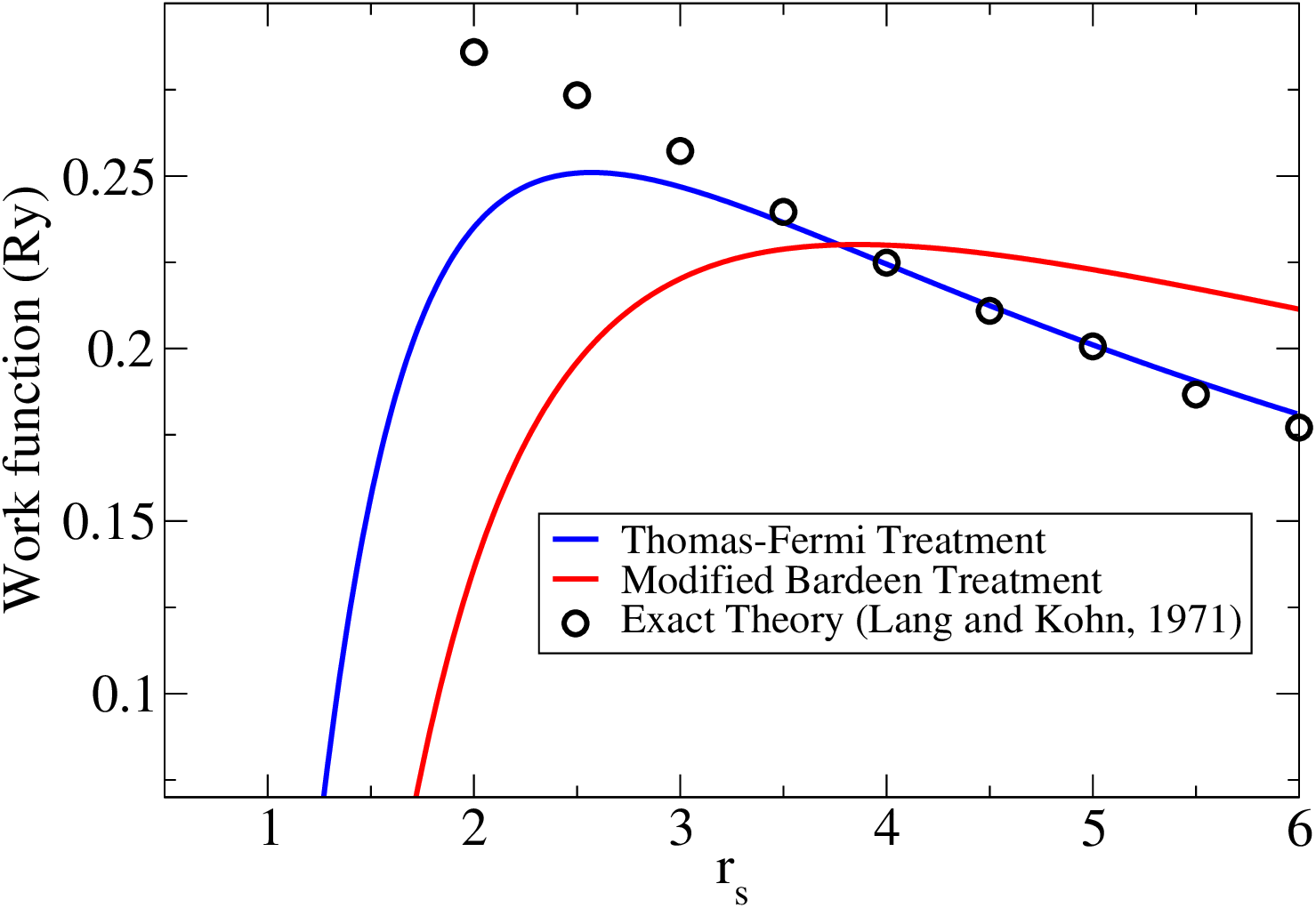}
\caption{Workfunction of a jellium metal at different densities:
blue curve - Eq.~(\protect{\ref{phis}}), 
red curve - Eq.~(\protect{\ref{phisqm}}),
circles - exact
theory \protect{\cite{lang1971a}}}
\label{workfunction7a}
\end{figure}

As a next step we integrate the Poisson's equation and employ
Eq.~(\ref{mahan}) to determine $\xi_0$ self-consistently. This leads
to a transcendental equation similar to Eq.~(\ref{phis}):
\begin{equation}
\label{phisqm}
\Phi_s\equiv\frac{3.68}{r_s^2}\xi_0=
\frac{1.627}{r_s}\bar f(\xi_0)-
\frac{2.21}{r_s^2}+\frac{0.916}{r_s}+
\frac{0.88}{r_s+7.8},
\end{equation}
where
\begin{equation}
\bar f(\xi_0) = {\phi_0}(\xi_0)+\frac{1}{2}\left[q_1^2(\xi_0)-q_2^2(\xi_0)\right],
\end{equation}
and
\begin{equation}
\phi_0(\xi_0) =
\frac{1+3\xi_0-3\xi_0\sqrt{1+\xi_0} 
{\coth}^{-1}(\sqrt{1+\xi_0 })}{4(1+ \xi_0 )}
\end{equation}
The value of $\Phi_s$ obtained from Eq.~(\ref{phisqm}) is shown in 
Fig.~\ref{workfunction7a} and compared with our previous Thomas-Fermi result.
We see that the Thomas-Fermi result is in better agreement with the numerical
calculations \cite{lang1971a}. We believe that this is due to the fact that
the total charge within the metal is better reproduced by a self-consistent
(but not oscillatory) Thomas-Fermi denisty rather than by an oscillatory
(but not fully self-consistent) denisty obtained from the modified 
Bardeen treatment.

\bibliographystyle{apsrev}
\bibliography{nsf_new}

\end{document}